\documentclass[aps,pra,floatfix,showpacs,noshowkeys,epsfig,graphics,natbib]{revtex4}
\usepackage{graphicx}
\usepackage{amsmath}
\usepackage{amsfonts}
\usepackage{amssymb}

\newcommand{\newc}{\newcommand}
\newc{\beq}    {\begin{equation}}
\newc{\eeq}    {\end{equation}}

\newc{\beqa}    {\begin{eqnarray}}
\newc{\eeqa}    {\end{eqnarray}}
\newc{\bs}    {\section}
\newc{\no}    {\\ \nonumber}

\newc{\st}    {\stackrel}

\begin{document}
\title{ Is dark energy from cosmic Hawking radiation?}
\author{Jae-Weon Lee}\email{scikid@gmail.com}
\affiliation{School of Computational Sciences,
             Korea Institute for Advanced Study,
             207-43 Cheongnyangni 2-dong, Dongdaemun-gu, Seoul 130-012, Korea}
\affiliation{ Department of energy resources development,
Jungwon
 University,  5 dongburi, Goesan-eup, Goesan-gun Chungbuk Korea
367-805}
\author{Hyeong-Chan Kim}
\email{hyeongchan@gmail.com}
\affiliation{Center for Quantum Spacetime, Sogang University,
Seoul 121-742, Republic of Korea
}
\author{Jungjai Lee}
\email{jjlee@daejin.ac.kr}
\affiliation{Department of Physics, Daejin University, Pocheon, Gyeonggi 487-711, Korea}

\date{\today}

\begin{abstract}
It is suggested that dark energy is the energy of the Hawking radiation
from a cosmic horizon. Despite of its extremely low
Gibbons-Hawking temperature, this radiation could have the
appropriate magnitude $O(M_P^2 H^2)$ and the equation of state to
explain the observed cosmological data if there is a Planck scale
UV-cutoff, where $H$ is the Hubble parameter.
\end{abstract}

\maketitle
\section{Introduction}

In this paper, we  suggest that the dark energy is the energy of the
 Hawking radiation  from a cosmic horizon.
 The dark energy  having  positive energy density
$\rho_\Lambda$, negative pressure
 $p_\Lambda$ and the equation of state
$w_\Lambda\equiv p_\Lambda/\rho_\Lambda$ smaller than $-1/3$ seems to cause the
accelerating expansion of the universe which can be inferred by
 the observations of the  type Ia supernova  ~\citep{riess-1998-116,perlmutter-1999-517},
the cosmic microwave background radiation (CMBR)~\cite{wmap3} and
the large scale structures~\cite{SDSS1}.
 The exact origin of this mysterious
dark energy  is one of the hardest puzzles in modern physics and
astronomy~~\cite{CC}. The dark energy problem consists of three sub-problems;
why the dark energy density is so small, nonzero, and comparable
to the critical density at the present. Although, there are
already various models relying on materials such as
quintessence~\cite{PhysRevLett.80.1582,PhysRevD.37.3406},
$k$-essence~~\cite{PhysRevLett.85.4438},
phantom~~\cite{phantom,nojiri-2006-38}, Chaplygin
gas~~\cite{Chaplygin}, and quintom~\cite{quintom} among
many~\cite{wei-2007,Gough:2007cj}, no model has solved the
problems in a natural manner so far.
The holographic dark
energy (HDE) models ~\cite{li-2004-603,PhysRevLett.82.4971} are based on the
holographic principle saying the number of degrees of freedom
inside a volume is proportional to the surface area of the region~\cite{holography} rather than the volume.
The connection between
 dark energy and the holographic principle implies
that dark energy could be related to a cosmic horizon at which the
 principle applied. The energy density of HDE is usually $O(H^2 M_P^2)$, which is similar to the observed value, where
 $H$ is the Hubble parameter.
This form of dark energy  is also often found in
 many dark energy models based on the quantum  vacuum fluctuation
~\cite{Zeldovich,Gurzadyan:2000ku,Padmanabhan:2007xy,Mahajan:2006mw,Djorgovski:2006vn,0264-9381-16-1-011,
Barbachoux:2007qb,Gariel:2005ts,
padmanabhan-2005-22,agegraphic,Myung:2008pi}.
However, it is still uncertain why the dark energy should be related to a cosmic horizon in this form
and why  a horizon property is related to dark energy density at the center of the horizon.

Recently, we have suggested  a new idea  that dark energy is
related to the entanglement of the quantum vacuum
fluctuation~\cite{myDE} or the erasure of the quantum information
of the vacuum at the cosmic horizon~\cite{forget}. This model
predicts $\rho_\Lambda$ and $w_\Lambda$ consistent with the
observational data. According to the model, the entanglement of
virtual particles around the expanding cosmic event horizon
induces a kind of thermal energy which can be interpreted as dark
energy. (This idea is followed by related
works~\cite{gough-2007,Horvat:2007ic,Medved:2008vh}.)
 One
possible interpretation for this entanglement dark energy is
that dark energy is just  the energy related to the cosmic Hawking radiation or Unruh radiation ~\cite{newscientist,sciencenews}.
According to Hawking, in the black hole physics, the entropy of
Hawking radiation can be identified with the entanglement entropy
between the outgoing and incoming particles created at the black
hole horizon due to the quantum
fluctuation~\cite{hawking,PhysRevD.34.373}. One can imagine that a
similar thing happens at a cosmic horizon and dark energy is the
energy of this cosmic Hawking radiation.

In this paper, we  investigate in detail the possibility that the dark energy is the energy of
 Hawking radiation itself from the cosmic horizon.
 In Sec. II we discuss the quantum nature of this dark energy.
In Sec. III we show the relation between the entropy of the cosmic Hawking radiation and
dark energy.
In Sec. VI the predictions of our theory are compared with the observational data
and cosmological constraints.
Section V contains discussions.

 \section{Quantum nature of  Dark energy}
At a first glance, the idea linking dark energy to the Hawking
radiation may look counterintuitive, because the Gibbons-Hawking
temperature of the present cosmic event horizon $T_H\sim
O(H)\sim10^{-60} M_P\simeq 10^{-32} eV$ ~\cite{mycoin} is extremely
low and thermal radiation usually has a positive equation of
state unlike dark energy.
However, it is known~\cite{RevModPhys.57.1} that  Hawking
radiation and  ordinary thermal radiation such as CMBR are very different in their nature.
 The ordinary
radiation is composed of excited quanta of which existence usually
has an observer independent meaning, while the presence of the
Hawking radiation, being the transformed vacua, is  observer
dependent~\cite{PhysRevD.15.2738}. For example, an observer at
rest far from a black hole can see the Hawking radiation from the
horizon, while  another observer freely falling toward  the black
hole could not see the radiation. Similarly, the cosmic Hawking
radiation seen by one observer $O_A$ could be just
 the vacuum fluctuation for another observer $O_B$ sitting at a different position. (See Fig. 1.)
Instead, the observer $O_B$ could see his/her own Hawking radiation from his/her own horizon.

 The quantum field theoretic calculations of the energy density of the cosmic Hawking radiation
 support this idea too.
 It is well known that the cosmic Hawking radiation in de Sitter universe has a
  energy momentum tensor $T_{\mu\nu}=\rho  g_{\mu\nu}$ and hence
 the equation of state $-1$ like the cosmological constant~\cite{RevModPhys.57.1}
 which  is different from that of the thermal radiation.
Unfortunately, this Hawking radiation is known to have too small energy density
~\cite{RevModPhys.57.1}, \beq \rho=\langle T_{00} \rangle\sim
\langle \phi^2\rangle \sim \int^{\infty}_{} k^2 \phi_k^2 dk\sim H^4,
\eeq
to explain the observed dark energy density and, hence, it has
never been considered as a candidate for dark energy. This value
was obtained by calculating
 Bunch-Davies vacuum  energy density
for a scalar field $\phi$
in de Sitter universe~\cite{vilenkin}.
Here,  $\phi_k$ is the mode
with the momentum $k$ on the curved space-time, which contains $H$ in its expression.
However,  this value was obtained by using the renormalization
technique with an infinite UV cut-off and not adequate when there is a finite
UV cut-off $k_u$, as in our case.
It was shown in ~\cite{UV} that with a finite UV-cutoff $k_u$, the energy density above should be changed to
\beq
\rho\sim \int^{k_u}_{} k^2 \phi_k^2 dk\sim H^2 k_u^2.
\eeq
 We notice that if we choose a natural value,
 $M_P$, as the UV-cutoff, the energy density of Hawking radiation is not $O(H^4)$ but  $O(H^2 M_P^2)$, which is just of
 order of
the dark energy density observed!
Thus, for a De Sitter-like universe, we can expect that
 the cosmic Hawking radiation is a plausible candidate for dark energy.

\section{Dark energy and cosmic Hawking radiation}
Let us roughly  estimate this dark energy (Hawking radiation energy)
density by assuming
 that the change of Hawking radiation energy  is  given by the change of ``horizon energy" ~\cite{BoussoDesitter}
\beq
\label{eenergy}
dE_{\Lambda}= T_H dS_{\Lambda},
\eeq
where $T_H$ is the temperature of the horizon.
In \cite{myDE} we use this usual `first law-like equation' to define  the entanglement energy.
The relation between this energy and causal horizons is investigated by Jacobson~\cite{PhysRevLett.75.1260}.
We are supposing that the emission of the cosmic Hawking radiation actually represents the change of the vacuum energy or the horizon energy
 satisfying Eq. (\ref{eenergy}) as in black hole cases.

The Hawking radiation for an observer could be just the vacuum state
for another observer. In cosmology, the Hubble horizon, the event horizon, and the apparent horizon
are usually considered for the comic horizon. Let us denote the radius of these generic horizons with
$r$. For the cosmic horizons the Hawking temperature is usually given by
$T_H\sim 1/r$ ~\cite{Gong:2006ma,Cai:2008gw} and is of order the Hubble parameter $H$.
Thus, the typical dark energy  from the cosmic Hawking radiation
is $E_\Lambda= \int dE_\Lambda\sim \int T_H dS_\Lambda \sim M_P^2 r$ and
its energy density is $\rho_\Lambda = 3E_\Lambda/4\pi r^3 \sim M_P^2/r^2\sim M_P^2 H^2$
which is the value just required.
Here, inspired by
the holographic principle, the dark entropy $S_\Lambda$
is assumed to be proportional to the horizon area in Planck unit, i.e., $S_\Lambda\sim M^2_P r^2$.
This `area-law' for the entropy related to the horizon is typically observed in the entanglement
theory~\cite{Srednicki}, string/M theory~\cite{Maldacena:1997re} and the quantum gravity theory~\cite{holography}.
(One can say that the Hawking radiation has a `surface' property while  the CMBR has a `volume' property,
which is the key property allowing the Hawking radiation  to be a plausible dark energy candidate.)
Therefore, this kind of dark energy  generally  has the form of HDE and
could explain the present value of the observed  dark energy density.
Although the HDE model is  a promising alternative, it has its own difficulties~\cite{ageproblem,Instability,Li:2008zq}
 and the exact origin of the HDE itself has yet to be adequately justified
 in the field theoretic context.

\begin{figure}[hbtp]
\includegraphics[width=0.32\textwidth]{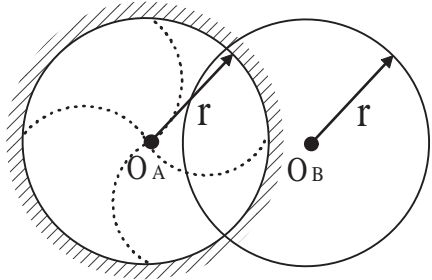}
\caption{
Hawking radiation (dotted lines) from a cosmic horizon with a radius $r$ can play a role of dark energy.
The radiation seen by one observer $O_A$
could  be
 the vacuum fluctuation for another observer $O_B$ sitting at a different position.
Instead, the observer $O_B$ could see his/her own Hawking
radiation from his/her own horizon.
}
\end{figure}

It is also worth mentioning that the cosmic Hawking radiation and Hawking radiation from a black hole
have  differences too. For the black hole, the Hawking temperature  increases
as the Hawking radiation is emitted, while for the universe the temperature decreases.
This may have a quantum informational origin~\cite{forget,myDE}.
In the cosmological case, the outside of the horizon is invisible to the observer,
while for the black hole the inside of the horizon is invisible.
We have seen that if the entropy  of the cosmic Hawking radiation satisfies the `area-law',
it is almost unavoidable  that the radiation behaves as dark energy.
The key issue now is how much entropy cosmic Hawking radiation actually has.

 How can we explicitly calculate the entropy of the cosmic Hawking radiation?
 Hawking et al. ~\cite{Hawking:2000da}, argued that the entropy of the de Sitter horizon can be described as the quantum entanglement
of the conformal field theory vacuum across the horizon and it
can also be viewed as the entropy of the thermal Rindler particles near the horizon, i.e., Hawking radiation.
In this paper we follow this interpretation.
They also showed that the entanglement entropy of four-dimensional space-time with a horizon such
as the de Sitter universe is
\beq
\label{Sents}
S_{Ent}=\frac{A_4}{4 G_4}=\frac{2 N_{dof} A_4}{a^2 \pi},
\eeq
where $a$ is a UV cut-off, $G_4$ is the 4-dimensional gravitational constant,
$N_{dof}$ is the degrees of
freedom,  and
$A_4$ is the area of the horizon. Thus, the 4-dimensional Planck mass is related to $G_4$ and the number of fields
in this theory.

Similarly, one can also calculate the entropy of the Hawking radiation using the concept of entanglement
of  quantum fields ~\cite{PhysRevD.52.4512,PhysRevLett.71.666}.
The entanglement entropy of a quantum field with a horizon is
generally expressed in the form
$S_{Ent}=\beta
r^2/a^2,$
  where $\beta$ is an $O(1)$ constant that
depends on the nature of the field.
Entanglement entropy for a single massless scalar field
 in curved backgrounds with a time-dependent event horizons
  is calculated in Ref. ~\cite{PhysRevD.52.4512,PhysRevLett.71.666} using a Hamiltonian approach.
By performing numerical calculations on a sphere lattice, $\beta=0.30$ was obtained
 for Friedmann-Robertson-Walker universe.
 If there are $N_{dof}$  spin degrees of
freedom  of  quantum fields  within  $r$, due to the additivity
of the entanglement entropy~\cite{nielsen}, we must add up
the contributions from all of the individual fields to
$S_{Ent}$~\cite{PhysRevD.52.4512}, that is, $S_{Ent}=N_{dof} \beta
r^2/a^2$. In this case dark energy becomes the entanglement dark energy in ~\cite{myDE}.

To explicitly calculate the dark energy density, let us choose a
specific horizon for the dark energy.
 It is well known that HDE with the particle horizon
  does not give an accelerating universe ~\cite{hsu}, while HDE with the event horizon
 does~\cite{li-2004-603}. Thus, henceforth,
 we investigate the most plausible case, where
  $r=R_h$ is the radius of the future event horizon
  \beq
\label{Rh}
R_h\equiv R(t)\int_t^\infty \frac{d t'}{R(t')}= R(t)\int_R^\infty \frac{d R(t')}{H(t') R(t')^2}.
\eeq
For the temperature $T_H$ of the Hawking radiation
 we use the Gibbons-Hawking temperature~\cite{PhysRevD.15.2738} $1/2 \pi R_h$
 and for the entropy $S_\Lambda$ we choose the entropy satisfying the holographic  principle, i.e,
 $S_\Lambda=\pi R_h^2 m_P^2$. This choice of $S_\Lambda$ corresponds to the first equality of Eq. (\ref{Sents}).
 This is a plausible choice because
our universe is going to a dark energy dominated universe,
which can be a quasi-de Sitter universe~\cite{0264-9381-6-6-014}.
 Here $m_P=\sqrt{8 \pi} M_P$  and
 we consider the flat ($k=0$) Friedmann universe described by the metric
$ds^2=-dt^2+R^2(t)d\Omega^2$.
By integrating Eq. (\ref{eenergy})
one can easily obtain  $E_\Lambda=8\pi R_h M_P^2$ and
~\cite{li-2004-603}
 \beq
 \label{holodark2}
\rho_\Lambda=\frac{3 E_\Lambda}{4\pi R_h^2}=\frac{3 d^2 M_P^2}{ R_h^2 }
\eeq
with the parameter $d=\sqrt{2}$.
This
$\rho_\Lambda$ decreases much slowly than $\rho_r$ and  could eventually dominate
the universe~\cite{mycoin}.

Alternatively, considering that the temperature is varying,
 one may use the minimum free energy  condition~\cite{forget} $dF=d(E_\Lambda-T_H S_\Lambda)=0$
 instead of Eq. (\ref{eenergy}), or,
\beq
\label{dE2}
dE_\Lambda=d(T_H S_\Lambda),
\eeq
which leads to $d=1$.
The current observational constraint is
  $d=0.91^{+0.26}_{-0.18}$~\cite{zhang-2007}, which is in a good agreement with our predictions.
Since our universe is not exactly equal to the de Sitter
universe, $S_\Lambda$ or  $T_H$ can be slightly different from that of the de
Sitter universe. Thus, it is expected that $d$ is $approximately$ 1
as observed.

\section{Observational constraints}
Let us move on the issue of the equation of state, or how the cosmic Hawking radiation could
accelerate the expansion of the universe.
Recall that the Hawking radiation,
even when it is the electromagnetic waves, does not have the  equation of state for thermal radiation, $1/3$.
It is important to note that, contrary to intuition, the (effective) equation of state of matter
in the universe is a dynamical property rather  than its intrinsic nature.
It depends  on how the energy density changes as the universe expands~\cite{myDE}.
(See Eq. (\ref{p}).)
For example a scalar field  rolling down very slowly its potential behaves
like dark energy (i.e., quintessence), while during a rapid oscillation around the potential
minima it  behaves like ordinary matter.

To be more precise, consider
 perfect  fluid
having  the energy momentum tensor
$T_{\mu\nu}=(\rho_\Lambda+p_\Lambda) U_\mu U_\nu- p_\Lambda g_{\mu\nu}$,
which the homogeneous dark energy should satisfies, where $U^{\mu}U_{\mu}=1$.
It is well known that, from the cosmological energy-momentum conservation equation $T^{\mu\nu}_{;\nu}=0$, one can obtain
\beq
\label{p}
p_\Lambda=\frac{d(R^3\rho_\Lambda)}{-3 R^2 dR }.
\eeq
This equation implies
that
the perfect fluid with increasing total energy within a comoving volume as the universe expands
has effective negative pressure and could play a role of dark energy~\cite{mycoin}.
When HDE dominates other matter density, its energy density
behaves as $\rho_\Lambda \sim R^{-2+2/d}$ ~\cite{li-2004-603}, hence,
the total dark energy in the comoving volume $\rho_\Lambda R^3$ is an increasing
function of time.
Thus, the Hawking radiation having the energy
in the form of HDE could act as dark energy  for $d>0$,
because  its energy density $\rho_\Lambda$ decreases
slower than $1/R^3$.

Interestingly, our theory can be verified even current observational data.
From Eq. (\ref{p})  one can  obtain the equation of state for
our dark energy in the form of HDE in Eq. (\ref{holodark2}) ~\cite{li-2004-603,1475-7516-2004-08-013}
\beq
\label{omega}
w^0_\Lambda =-\frac{1}{3} \left(1+\frac{2\sqrt{\Omega^0_\Lambda} }{d}\right),
\eeq
and
the change rate of it at the present time ~\cite{li-2004-603,1475-7516-2004-08-006}
\beqa
\label{omega3}
w_1&\equiv& \left.\frac{d w_{\Lambda}(z)}{dz}\right|_{z=0}\no
 &=&
\frac{\sqrt{\Omega^0_\Lambda} \left( 1 - \Omega^0_\Lambda  \right)}{3d} \left(1+\frac{2\sqrt{\Omega^0_\Lambda} }{d}\right)
\eeqa
where  $z$ is the redshift parameter,
 $w_\Lambda (z) \simeq w^0_\Lambda+w_1 (1-R)$, and
   the observed present  value of
the density parameter
of the dark energy $\Omega^0_\Lambda\simeq 0.73$.
For $d=1$ these equations give $w^0_\Lambda=-0.903$ and $w_1=0.208$, while
for $d=\sqrt{2}$,  $w^0_\Lambda=-0.736$ and $w_1=0.12$.
The predictions of our theory well agree with the recent observational data;
$w^0_\Lambda = -1.03 \pm 0.15$ and  $w_1 = 0.405^{+0.562}_{-0.587}$ ~\cite{xia-2007,Zhao:2006qg}.
If we interpret the entanglement dark energy in \cite{myDE} as the energy of the cosmic Hawking radiation,
then $w^0_\Lambda = -0.93$ and  $w_1 = 0.11$.
(See Fig. 2.)
Thus, the cosmic Hawking radiation can give the appropriate equation of state for the observed dark energy, because
the Hawking radiation gives the HDE with $\rho_\Lambda$ in Eq. (\ref{holodark2}).

\begin{figure}[hbtp]
\includegraphics[width=0.4\textwidth]{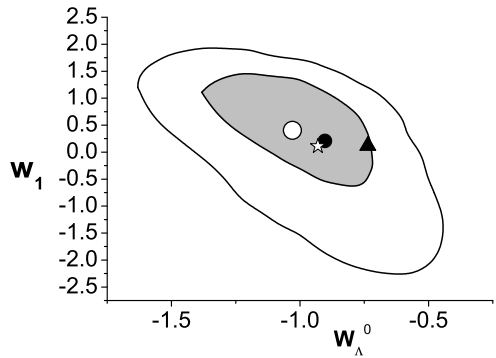}
\caption{Observational constraints on the equation of state
parameters $w^0_\Lambda$ and $w_1$ (Fig. 2 of \cite{xia-2007})
and our theoretical predictions.
 The contours represent 95\% and  68\% C.L.,
respectively. The white dot denotes the best-fit point from the
observations. The black dot represents our theoretical prediction
with $dE_{\Lambda}= d(T_H S_{\Lambda})$ and  the triangle with
$dE_{\Lambda}= T_H dS_{\Lambda}$.
The white star indicates the results for the entanglement dark energy in \cite{myDE}.}
\end{figure}

 It was also shown that the dark energy in this form
 can solve the cosmic coincidence problem too,
  if there was an ordinary inflation with the number of e-fold $N\simeq 65$ at the very early universe~\cite{mycoin,li-2004-603}.
As is well known, to solve many problems of the standard big-bang
cosmology we need an inflation  $N\st{>}{\sim} 60$. Note that
there is also a conjectured theoretical upper bound, $N
\st{<}{\sim} 65$ from the holographic principle
~\cite{Banks:2003pt,wang:104014,Kaloper:2004gp} for the
asymptotic de Sitter universe or from the density perturbation
generation ~\cite{PhysRevLett.91.131301,PhysRevD.68.103503}.
Thus, very interestingly, the required $N$ value falls within the
narrow range expected from other cosmological constraints.
It is also interesting that the HDE model is connected to the second law of thermodynamics~\cite{second}.

Now, we investigate whether the cosmic Hawking radiation is consistent with other cosmological constraints.
The typical (peak) wavelength of the cosmic Hawking radiation at present is  $\lambda\sim 1/T_H\sim1/H$,
which is comparable to the size of the observable universe. This
huge wavelength explains why we could not detect dark energy
directly so far and why dark energy seems to be so homogenous.
 For example, let us consider the decoupling of
electron from the cosmic Hawking radiation. As is well known the
electron is decoupled from the radiation when the scattering rate
$\Gamma=n_e \sigma_T$ is smaller than the Hubble parameter $H$.
Here, $n_e$ is the free electron number density, and
$\sigma_T=6.65 \times 10^{-25} cm^2$ is the Thomson cross section
~\cite{kolb90}. Since the binding energy of hydrogen, $13.6~eV$,
is much larger than $T_H$, the ionization of hydrogen and the
increase of $n_e$ due to the Hawking
 radiation is negligible and hence we can ignore the electromagnetic effect of the Hawking radiation on the
 hydrogen as long as $T_H \ll 13.6~eV$.
 Another example is the effect of photons in the Hawking radiation on the ultra high energy cosmic ray (UHECR, mainly proton).
 In this case we can obtain Greisen-Zatsepin-Kuzmin (GZK)-like  cut-off ~\cite{GZK} for the  energy of cosmic rays
 interacting with the Hawking radiation
 ~\cite{stanev,GZKlimit,GZK2}
 \beq
 E_p < \frac{m_\pi}{4 T_H} (2 m_p+m_\pi)\simeq  6.7\times 10^{39}~GeV,
 \eeq
 where, we used $T_H=10^{-32} eV$ for the photon energy instead of
 the energy of usual CMBR photon, and $m_p$ is the proton mass and $m_\pi$ is the pion mass.
 This bound is well above the Planck energy and the observed constraints
 $E_p < 6\times 10^{10}~GeV$ ~\cite{GZKobservation}. Thus, the effect of the Hawking radiation photons on the UHECR is negligible compared to that of CMBR too.
 Furthermore, the temperature of the cosmic Hawking radiation is too low to distort
   the CMBR signal ~\cite{Beck:2004fh}.

The cosmological principle implies
the homogeneity of the Hawking radiation  and this also justifies the
simple-minded use of the spherical volume $4\pi R_h^3/3$ during
the above calculation of $\rho_\Lambda$ at the center of the
horizon sphere.
Hawking radiation also provides a mechanism linking
the local dark energy density to the global property of
the horizon, because the gravitational field at the center plays a role of a detector
for the energy density of the radiation from the horizon.

This dark energy neither spoils the standard inflation scenario.
Note  that this dark energy  is different from the energy  from quantum fluctuation of an inflation
field $\phi$, even though the both energy is related to the quantum fluctuation of the vacuum.
The inflaton energy density fluctuation comes from the fluctuation of the arrival time for the inflaton
to roll down to the potential minima
due to the quantum fluctuation $\delta \phi$ of the inflaton field. Thus, it is highly dependent on the
shape of the potential. On the other hand, the energy of Hawking radiation is
from all quantum fields regardless of its potential.
Furthermore, in \cite{mycoin}, it is shown that the size of the event horizon is an exponentially
increasing function of time $t$ during the inflation, i.e.,
\beq
R_h(t)\simeq \frac{1}{H_i}(1+A e^{H_i(t-t_i)})
\eeq
where A and $t_i$ are  constants and $H_i$ is the
 Hubble parameter at the inflation. Thus,  after some e-folds of expansion, the energy density
  of Hawking radiation which is proportional to $1/R_h^2$
 is negligible compared to the inflaton energy density and does not spoil the usual
 scenario of inflation.

Although, $\rho_\Lambda$ has no $\hbar$ explicitly,
the quantumness of the Hawking radiation  can be seen clearly from the existence of $\hbar$ in Gibbons-Hawking temperature
$T_H=\hbar/ 2\pi R_h$, where $R_h$ is the radius of the cosmic event horizon.
 Does the quantum nature of Hawking radiation spoil the classical description of back ground geometry?
To see this is not the case, consider the back-reaction of the quantum fluctuation in chaotic inflation
as an example.
This is a second order effect in the cosmological perturbation.
The fractional contribution
of scalar metric perturbations $\rho_s$ to the total energy density $\rho_0$ is shown ~\cite{PhysRevD.56.3248} to be
\beq
\frac{\rho_s }{\rho_0} \sim -\frac{m^2 \phi^2}{ M_P^4},
\eeq
which is negligible when $\phi\st{<}{\sim} M_P$ and the inflaton mass $m\ll M_P$.
The energy scale of dark energy today  is even much lower than this inflation scale. Furthermore,
to make the local space-time fluctuate significantly due to the quantum  effect
the energy density of dominant matter should be at least $O(M_P^4)$~\cite{Linde:2005ht}.
 Thus, generally, the quantum fluctuation of the space-time due to the Hawking radiation
  is small compared to the quantum gravitational scale, and
 we can treat the evolution of the universe classically
 even when the dark energy from
 Hawking radiation dominates the universe.
The cosmological effect of the Hawking radiation is relevant only at the cosmological scale
in an average sense.

\section{Discussion}
We can summarize how the cosmic Hawking radiation could solve the dark energy problem.
In this scenario,
the dark energy density is small due to the holographic principle,
 comparable to the current critical density because the number of e-folds
 during the inflation $N\simeq 65$
 or the horizon size is  $O(1/H)$,
and non-zero due to the unavoidable quantum fluctuation.
The cosmic Hawking radiation  also explains why dark energy
is so homogeneous and can hardly  interact with ordinary matter.
Its wavelength is simply too long.

Let us discuss the possible  direction of future research.
Although the event horizon is the most natural candidate
for our purpose from the viewpoint of quantum information science and  gravity,
it is still possible that  the proper
horizon for dark energy could be other horizon  besides the event horizon ~\cite{Melia}
 such as the apparent horizon \cite{1475-7516-2007-01-024}.
In this case the interaction between dark matter and Hawking radiation
would be important.
Since $\omega^0_\Lambda$ and $\omega_1$
will be precisely constrained by observations in the near future,
our model can be verified soon, once we derive the properties of the cosmic Hawking radiation
more precisely. Thus, deriving the exact entropy or energy of the cosmic Hawking radiation would be
of great importance
in this direction.
 The temperature of the cosmic Hawking radiation may be too low to be detected
 above the CMBR ~\cite{Beck:2004fh}, however,
 the cosmic Hawking radiation  and its energy
 may be simulated using the acoustic horizons ~\cite{PhysRevLett.46.1351,PhysRevD.58.064021} or optical black holes~\cite{opticalBH} in laboratories.
 If dark energy is really the energy of the cosmic Hawking radiation, this model provides
 not only a new direction to solve  the dark energy problem, but also the first observational evidence
 for the Hawking radiation itself.

\section*{acknowledgments}
This work was supported in part by the topical research
program (2009 -T-1) of Asia Pacific Center for
Theoretical Physics.


\end{document}